# The Renaissance of Black Phosphorus

*Xi Ling[1], Han Wang[2]\*, Shengxi Huang[1], Fengnian Xia[3], Mildred Dresselhaus[1,4]\**

[1]Department of Electrical Engineering and Computer Science, Massachusetts Institute of Technology, Cambridge, Massachusetts 02139, USA
[2]Ming Hsieh Department of Electrical Engineering, University of Southern California, Los Angeles, California 90089, USA
[3]Department of Electrical Engineering, Yale University, New Haven, Connecticut 06511, USA
[4]Department of Physics, Massachusetts Institute of Technology, Cambridge, Massachusetts 02139, USA

\***Corresponding Authors:**

**Prof. Mildred S. Dresselhaus**

**Department of Physics and Electrical Engineering and Computer Science,**

**Massachusetts Institute of Technology, Cambridge, Massachusetts 02139, USA**

**Tel: +1-617-253-6864**

**Email:** millie@mgm.mit.edu

**Prof. Han Wang**

**Ming Hsieh Department of Electrical Engineering,**

**University of Southern California, Los Angeles, California 90089, USA**

**Tel: +1-213-821-4293**

**Email:** han.wang.4@usc.edu

**Abstract**

One hundred years after its first successful synthesis in the bulk form in 1914, black phosphorus (black P) was recently rediscovered from the perspective of a two-dimensional (2D) layered material, attracting tremendous interest from condensed matter physicists, chemists, semiconductor device engineers and material scientists. Similar to graphite and transition metal dichalcogenides (TMDs), black P has a layered structure but with a unique puckered single layer geometry. Because the direct electronic band gap of thin film black P can be varied from 0.3 to around 2 eV, depending on its film thickness, and because of its high carrier mobility and anisotropic in-plane properties, black P is promising for novel applications in nanoelectronics and nanophotonics different from graphene and TMDs. Black P as a nanomaterial has already attracted much attention from researchers within the past year. Here, we offer our opinions on this emerging material with the goal of motivating and inspiring fellow researchers in the 2D materials community and the broad readership of *PNAS* to discuss and contribute to this exciting new field. We also give our perspectives on future 2D and thin film black P research directions, aiming to assist researchers coming from a variety of disciplines who are desirous of working in this exciting research field.

**Introduction**

At the beginning of 2014, a few research teams including the ones led by the authors reintroduced black phosphorus (black P) from the perspective of a layered thin film material (1–6), in which new properties and applications have arisen. Since then, black phosphorus, the most stable allotrope of the phosphorus element, is emerging as a promising semiconductor with a moderate band gap for nanoelectronics and nanophotonics applications (7, 8). Its single- and few-atomic layer forms can be isolated by techniques such as micromechanical exfoliation, giving rise to a novel type of two-dimensional (2D) material with many unique properties not found in other members of the 2D materials family. Here, we present our perspectives on this latest addition to the 2D material family, which can bridge the energy gap between that of graphene and transition metal dichalcogenides (TMDs), such as molybdenum disulfide (MoS$_2$), molybdenum diselenide (MoSe$_2$), tungsten disulfide (WS$_2$) and tungsten diselenide (WSe$_2$). In addition, we also offer our viewpoint on utilizing the in-plane anisotropy of black P to develop conceptually novel electronic, photonic, and thermoelectric devices.

Black P is a single-elemental layered crystalline material consisting of only phosphorus atoms (9). Unlike in group IV elemental layered materials, such as graphene, silicene or germanene, each phosphorus atom has five outer shell electrons. Black P has three crystalline structures (10): orthorhombic, simple cubic, and rhombohedral. Semiconducting puckered orthorhombic black P is of interest here and it belongs to the $D_{2h}^{18}$ point group (see Fig. 1(a) and 1(b)), which has reduced symmetry compared to its many group IV counterparts (such as graphene) having the $D_{6h}^{4}$ point group symmetry. The single layer black P includes two atomic layers and two kinds of P-P bonds. The shorter bond length of 0.2224 nm connects nearest P atoms in the same plane, and the longer bond length of 0.2244 nm connects P atoms between the top and bottom of a single layer. The top view of black P along the z direction shows a hexagonal structure with bond angles of 96.3° and 102.1° (11, 12).

Early work on black phosphorus can be dated back to the first decade of the last century. Bridgman et al. (13) successfully obtained black phosphorus for the first time in 1914 by conversion from white phosphorus at a pressure of 1.2 GPa and an elevated temperature of 200 $^0$C. Unlike the white form of the phosphorus allotrope, black phosphorus is stable at ordinary temperatures and pressures. Bridgman was awarded the Nobel Prize in 1946 for "the invention of an apparatus to produce extremely high pressures, as well as the discoveries he made therewith in the field of high pressure physics". However, at that time, there was not much interest in the black P material itself. Furthermore, research on black P has been relatively silent for 100 years. There are only about 100 publications in the past 100 years on black P to our knowledge. Nevertheless, the research on black P has made steady progress in the decades that followed, including the study of its structural (14, 15), transport (16), optical, phonon (17), (18), and superconducting (19, 20) properties, as well as applications in battery electrodes (21–23). In 1953, Keyes (24) studied the electrical properties of black P. It is worth noting that a significant portion of the work in the 1970s and 1980s was performed by a few Japanese groups that made important progress in black P research, including key studies on its electrical (16, 25, 26) and optical properties (27–29), and the successful n-type doping of black P by tellurium (30).

These early studies of black P as a bulk material, however, did not catch much attention from the semiconductor research community at that time, likely due to the dominant role of silicon. Only since 2014, building on the study of graphene, few-layered hexagonal boron nitride (hBN) and TMDs in the last decade, black P has been rediscovered from the perspective of a 2D and thin film material. As a result, the recent surge of black P research since early 2014 has mainly focused on the material in its single-layer, few-layer or thin film forms where new properties have arisen and novel applications may be developed. Within one year, more than 70 papers were published on black P thin film, both in theory and experiment, covering the topics from band structure (31–33), strain (5, 34–38), defects (39), intercalation (40), and structure varieties (blue phosphorus (41), phosphorus nanotube (42), phosphorus nanoribbon (43–47), and stacked bilayer phosphorus (48)), to characterization methods (49–51), stability and passivation methods (52–55), novel physics properties (56, 57), and promising applications in electronics (1–4, 58–61)**,** photonics (62–67), thermoelectrics (68, 69) and gas sensing (70) devices. In addition, the hybrid structures between black P and other 2D materials were also studied for optoelectronics applications (45, 65) and its strong in-plane anisotropy brings new opportunities to inventing conceptually new electronic and photonic devices (6, 35, 71).

Recent theoretical studies have predicted that monolayer BP can have an extremely high hole mobility (10,000 cm$^2$V$^{-1}$ s$^{-1}$)(6). Besides the thickness of black P, strain is predicted to be an effective way for tuning the band gap of black P (5, 34–38). Rodin et al. (5) reported that uniaxial compressive strain can switch black P from nearly a direct band gap semiconductor to an indirect band gap semiconductor, semimetal, and metal. The modification of the energy band structure can be even richer in a black P nanoribbon by controlling the edge to be zigzag or armchair, as well as the functional groups at the edges (46). In addition, new physics, such as a negative Poisson's ratio (56), and a quasi-flat edge band (72) were reported due to the anisotropic honeycomb lattice. Also a giant Stark effect was predicted in nonchiral phosphorene nanoribbons (73). The potential of the material for high performance thermoelectric devices (69) and superior gas sensing (70) were also explored from a theoretical perspective. Those predictions indicate that black P is a promising candidate for many of these applications.

*Bridging the Energy Gap*

For many important applications in nanoelectronics and nanophotonics, the moderate band gap of black P (~ 0.3 eV) in its thin film form (thickness > 4 nm, or 8 layers) can bridge the energy gap between the zero-bandgap of graphene (74, 75) and the relatively large band gap of many transition metal dichalcogenides (1.5 eV-2.5 eV) (76–79) (Fig. 2(a)). The energy band structure of bulk black P obtained using angle-resolved photoemission spectroscopy (ARPES) is shown in Fig. 1(c). Recent studies have experimentally demonstrated the strong optical conductivity of back P thin film in the 1-5 μm wavelength range, revealing black P as an appealing candidate for near and mid-infrared optoelectronics as detectors, modulators and potentially as light generation devices like light emitting diodes (LED) and lasers. Recently, detectors (64) and imaging (62) functions have been demonstrated using black P thin films. A more attractive feature of black P for optoelectronics applications is the wide tuning range of its band gap with layer number (80) and with the application of strain (5). Several groups have theoretically predicted the quasi-particle band gap tunability in single- and few-layer black P, estimating that it can vary from 0.3 eV in bulk form to above 2.0 eV in its single-layer form (5), (26), as shown in Fig.1(d). This was further confirmed using infrared relative extinction spectra and scanning tunneling microscopy (STM) measurements. As shown in Fig. 3(a), Xia et al. (2) observed the absorption peak at around 2700 cm$^{-1}$ (~0.3 eV), which originated from the band gap of black P. In Fig. 3(b), the d$I$/d$V$ curve from STM characterization on black P measured by Liang et al. (81) indicates that the electronic band gap of single-layer black P is 2.05 eV. In addition, compressive and tensile strain can lead to a significant modification of the black P band structure especially in its single- and few-layer form (34, 72). As a result, single layer to thin film black P can cover a very broad energy spectrum and interact strongly with electromagnetic waves in the mid-infrared, near-infrared and visible frequency range where many important applications in defense, medicine and communication lie, such as night vision, thermal imaging and optical communication networks.

*Bridging the Gap in the Mobility/On-Off Ratio Spectrum*

The transport properties of black phosphorus lie between that of graphene and most TMDs previously studied. Fig. 2(b) shows the "mobility/on-off ratio spectrum" where we have plotted the mobility of the material in relation to the on-off current ratio of transistors enabled by them. Despite the possible variations of the mobility at different device operational conditions, transistor devices based on different 2D materials in general fall into different zones in the "mobility/on-off ratio spectrum" as shown in Fig. 2(b). Each region of this spectrum corresponds to some key application domains in nanoelectronics. Graphene is a 2D semi-metal with very high mobility, but the on-off ratio of graphene transistors is often less than 10 due to its zero-bandgap. On the other hand, many monolayer TMD materials have lately attracted much attention. The carrier mobility is usually relatively low (mostly lower than 100 cm$^2$/V.s) in these materials, but the on-off ratio of their transistors is very high, being easily above $10^8$ and may reach $10^{10}$ in some cases. TMD materials are hence appealing for ultra-low-power nanoelectronics. The mobility/on-off ratio combination for BP falls into a region on the plot not easily covered by graphene nor transition metal dichalcogenides such as MoS$_2$. This is a region where the mobility of the material is in a range of a few hundred to over 1000 cm$^2$/V.s and at the same time the on-off ratio of the device needs to be in the range of roughly around 10$^3$ to 10$^5$. Such properties of BP may be attractive for building GHz frequency thin film electronics. L. Li et al. measured a Hall mobility of around 210 cm$^2$/V.s at room temperature and above 350 cm$^2$/V.s along a randomly chosen direction in an 8 nm thick black P sample (1). Xia et al. measured their Hall mobility along the x-direction of a 15 nm thick black P thin film above 600 cm$^2$/V.s at room temperature and above 1000 cm$^2$/V.s below 120 K (Fig. 4(b)) (2). Field-effect mobilities in a similar range were also reported by various groups (Fig. 4(a)) (1, 4, 55, 63–65). Along the x-direction in bulk black P, the Hall mobility of holes exceeds 1000 cm$^2$/V·s at 300 K and 55,000 cm$^2$/V·s at 30 K, respectively. The electron mobility along the x-direction is also close to 1000 cm$^2$/V·s at 300 K and is above 10,000 cm$^2$/V·s at 50 K (10). These features are critical for building transistors with high current and power gains that are the most important attributes for constructing high frequency power amplifiers and high speed logic circuits. In addition, transistors based on black P thin film showed excellent current saturation and on-off current ratio above 10$^5$ (Fig. 4(a)) (1), both offering key advantages over graphene transistors for analog and digital electronics. Some detailed discussion on the electrical contact (59, 82) and effects of dielectric capping (60) have also been reported. Al$_2$O$_3$ overlayers were effectively utilized to protect black P devices for better stability and reliability, as well as to reduce the noise level of the transistors (53, 55). Recently, Wang et al. (58) demonstrated the operation of black P field-effect transistors at gigahertz frequency for the first time (Fig. 4(c-d)). The standard ground-signal-ground (GSG) pads were fabricated to realize the transition from microwave coax cable to on-chip coplanar waveguide electrodes. It shows that the short-circuit current-gain cutoff frequency $f_T$ is 12 GHz and the maximum oscillation frequency $f_{max}$ is 20 GHz in 300 nm channel length devices (Fig. 4(d)). Comparing to the graphene transistors, these first-generation high-speed black P transistors already show its superior performance for Radio-frequency (RF) electronics in terms of voltage and power gain due to the good current saturation properties arising from the finite black P band gap. Therefore, black P is a promising candidate for future high performance thin film electronics technology for operation in the multi-GHz frequency range and beyond.

*In-plane Anisotropy for Novel Device Applications*

While black P may well offer promising advantages over graphene and TMDCs in many traditional domains of nanoelectronics and nanophotonics, the most exciting application of black P may yet arise from its unique properties - the in-plane anisotropy (2, 6)that generates opportunities for designing conceptually new device and applications. With its puckered orthorhombic structure of the $D_{2h}$ point group, the effective mass of carriers of black P along the zigzag direction is about ten times larger than that along the armchair direction (16), which induces strong in-plane anisotropy in its electronic (6), optical (2, 6) and phonon properties (2). Such properties are shared by other lesser-known layered TMDs such as rhenium disulfide ($ReS_2$) and rhenium diselenide ($ReSe_2$), and together they may enable a new domain of electronics and photonics device research where the strong anisotropic properties of 2D materials can be used to invent new electronic and optoelectronic device applications. Here, we introduce two possible examples: (i) plasmonic devices with intrinsic anisotropy in their resonance properties, and (ii) high efficiency thermoelectrics utilizing the orthogonality in the heat and electron transport directions.

Recently, Low et al.(83) reported in *Physical Review Letters* theoretical work predicting the anisotropic plasmon resonance properties in black P atomic crystals, as shown in Fig. 5(a). In graphene plasmonic devices with disk geometry, the plasmon resonance frequency only possesses a scalar dependence on the momentum wave vector *q* defined by the size of the disks. In clear contrast, the collective electronic excitations in black P exhibit a strong in-plane anisotropy. The plasmon resonance in black P devices will have a vectorial dependence on the momentum. So simply by changing the linear polarization direction of the incident light, the plasmon resonance frequency of the structure can be continuously tuned (Fig. 5(b)). The tuning range will depend on the level of anisotropy in its x- and y- direction conductivities, the dielectric environment and the specific pattern design. This gives plasmonic devices even with highly symmetrical geometry (such as disks) an additional tuning knob - the light polarization that is unavailable in conventional metal-based plasmonics and graphene plasmonic devices where the material properties are largely isotropic.

Thermoelectrics is another field where the anisotropic transport properties of nanomaterials may enable significant performance improvements. Thermoelectric devices rely on the Seebeck effect to convert heat flow into electrical energy. Such devices will have many applications in developing solid-state, passively-powered portable electronic systems. The conversion efficiency is proportional to the ratio of a device's electrical conductance to its thermal conductance, which is collectively quantified by the thermoelectric figure of merit (ZT). It is highly desirable to achieve high electrical and low thermal conductivities simultaneously, in order to maximize ZT. In a paper recently published on *Nano Letters* (69), first-principles calculations revealed that monolayer black P exhibits spatially-anisotropic electrical and thermal conductances. Because the prominent electronic transport direction (armchair) is orthogonal to the prominent heat transport direction (zigzag), the ratio of these conductances can be significantly enhanced (Fig. 5(b)). It is predicted that ZT in monolayer black P can reach 2.5, which will meet the requirements for commercial use, along its armchair direction at 500 K**.** ZT is also greater than 1 at room temperature with moderate doping ($\sim 2\times 10^{12}$ cm$^{-2}$). Hence, black P is a mechanically flexible material that can naturally allow high efficiency heat energy conversion at room temperatures ($\sim$ 300 K) without any complex engineering. Moreover, Lv et al. (84) also addressed the large thermoelectric power factors in black P under an optimal doping level. Zhang et al. (85) reported that the semiconducting armchair phosphorene nanoribbon is a promising candidate for thermoelectric applications. These varieties of research opportunities offer extensive exploration space to experimentalists.

To study those fascinating properties and to achieve new applications based on its anisotropic structure, a reliable method to quickly and nondestructive identify the crystal orientation of a black P sample is urgently needed. IR spectroscopy was successfully used by the Yale/IBM team to identify the crystal orientation of black P samples, several tens of micrometers in size (2). The IR absorption along the armchair direction reaches a maximum due to the anisotropic absorption of black P, as shown in Fig. 3(a). Raman spectroscopy is generally considered as a fast and nondestructive method for materials characterization, and is effective for flake sizes down to a few micrometers or even smaller. The three typical Raman modes in black P, $A_g^1$, $B_{2g}$, and $A_g^2$, were reported to have different laser polarization dependences, which are strongly related with the crystal orientation (86). The spectroscopy feature in this unique confined and in-plane anisotropic structure itself is an interesting topic for exploring new physics in black P. Recently, X. Wang et al. (71) posted a study on *arXiv* reporting the highly anisotropic and tightly bound excitons in black P using polarization-resolved photoluminescence measurements. The exciton binding energy was extracted from the energy difference between the excitonic emission peak and the quasi-particle band gap, which is found to be as large as 0.9±0.1 eV. The results indicate that the electron, phonon, exciton and other many-body effects in black P are full of novelties and spectroscopy techniques are likely to play a critical role in future studies.

**Large scale synthesis and materials stability**

The future success of black phosphorus in electronics and photonics applications will critically hinge upon the development of reliable large-scale synthesis methods. Synthesis of black P can be traced back to one hundred years ago. In 1914 Bridgman (13) first reported a method to convert white P to black P at a moderate temperature of 200 °C and high pressure of 1.2 GPa within 5 to 30 min, while recently in 2012 Rissi et al. (87) reported that amorphous red P could be transformed into crystalline black P at $7.5 \pm 0.5$ GPa even at room temperature. By melting black P at a temperature of 900 °C and under a pressure of 1 GPa, black P single crystals larger than $5 \times 5 \times 10$ mm$^3$ can be achieved, as reported by Endo et al. in 1982 (88). Alternative techniques without using high pressure have also been developed, such as the technique involving mercury as a catalyst, developed by Krebs et al. in 1955 (89), the bismuth-flux based method by Brown et al. in 1965 (15), and the method based on a chemical transport reaction by Lange et al. in 2007 (90) that can utilize a relatively simple setup while avoiding toxic catalysts or "dirty" flux methods (91, 92). However, to the best of our knowledge, all the methods developed so far focused on the synthesis of bulk black P crystals but not its thin film or 2D forms at a wafer scale. This is most likely due to the fact that few have ever considered black P from the perspective of a 2D material before the recent revival of interest in this material. In future research, more effort combining expertise in materials science and chemistry should be devoted to the development of large-scale synthesis method for black P thin film or single- and few-layer nanosheets at the wafer scale where more application opportunities lie. It is also important to develop methods that can synthesize large-area single crystal thin films in which the anisotropic properties of black P may be explored at larger scales.

While bulk crystals of black P are stable under ambient conditions for at least a few months, black P in its single- and few-layer forms are found to be unstable in the presence of the moisture and oxygen in air (49). Samples of 10 nm thickness without proper protection can degrade in days while single-layer and few-layer samples may degrade within hours. Mark Hersam's group at Northwestern University reported their detailed X-ray photoelectron spectroscopy (XPS), atomic force microscopy (AFM), and Fourier-transform infrared spectroscopy (FTIR) characterizations that elucidate its underlying degradation mechanisms (53). XPS characterization shows that the $PO_x$ peaks appeared after exposing black P to air for one day, and FTIR characterization has also observed the P-O stretching mode at around 880 and P=O stretching mode at around 1200 cm$^{-1}$, suggesting the formation of oxidized phosphorus species that lead to the degradation of the material. However, after being encapsulated by the $Al_2O_3$ overlayers, the black P flakes are stable for at least several weeks in an ambient environment (Fig. 6(a)). Moreover, other teams have reported the faster degradation of black P in air (52, 54). Farvon et al. recently posted their study on *arXiv* (54), showing the photo-oxidation of black P exposed to laser light by in-situ Raman and transmission electron spectroscopic characterization. The oxidation rate is predicted to depend exponentially on the square of the energy gap of the layer. At this point, developing effective protection methods to slow down and eliminate the degradation process are needed. Several recent experiments demonstrated the use of oxidized aluminum as a passivation layer to isolate the black P surface from the ambient (53, 93), which works effectively in reducing the degradation of a relatively thick sample (~5 nm). Other techniques, such as PMMA coating (4), graphene and h-BN encapsulation (Fig. 6(b)) (94) have also been proposed for the same purpose with various levels of success. Overall, black P has good intrinsic thermal stability and the material is stable at high temperature if isolated from water and oxygen. Black P might not be as stable as other 2D materials, such as graphene and TMDs, in the presence of oxygen and water, but there are already breakthroughs in developing effective passivation methods to overcome this degradation issue. Learning from the commercial success of relatively unstable materials like organic semiconductors, and the technological importance of many toxic and potentially unstable materials like mercury cadmium telluride (HgCdTe), we believe that the stability issue should not be viewed as a show stopper preventing further research on this material. It is most likely that good passivation and packaging technology can resolve this issue. In fact, passivation and packaging are essential even for many of the commercialized semiconductors, such as silicon and III-V materials, to allow better device reliability and performance. There are many such techniques used by the semiconductor industry that we can also apply to protect black P devices, and such studies can constitute an interesting and important direction for future research.

*Future Directions*

In summary, we have already seen some interesting, but sporadic, research activities since early 2014 demonstrating black P-based detectors, modulators, RF transistors, sensors, etc., but both the fundamental study and application research on layered black P are still in their infancy with many unresolved issues and unexplored ideas. Here, we discuss a few topics for future research of black P that may be of interest to the research community in general. On the fundamental side, it will be very interesting to study the behavior of various types of polaritons and their dependence on the crystal orientation in single- and few-layer black P, such as plasmon and exciton polaritons. Advanced transport characterizations, such as angle-resolved quantum Hall effect in single- or few-layer black P, are important research topics for understanding the carrier dynamics of this material in the limit of 2D quantum confinement, subject to strong in-plane anisotropy. For nanoelectronics applications, thin film black P with thicknesses in the range of 4-10 nm may offer the best trade-off between mobility and on-off current ratio that is very attractive for developing high-speed flexible electronics systems that can operate in the multi-GHz frequency range and beyond. As a semiconductor with a respectable mobility and a moderate band gap, both analog and digital electronics can be constructed based on black P. With the availability of both p-type and n-type (doped in a controlled manner using tellurium) black P crystals, complementary metal–oxide–semiconductor (CMOS) circuit configurations may be realized using black P alone. For photonics application, black P is the most suitable for optoelectronic devices in the mid- and near-infrared spectrum ranges. By controlling the layer number and strain, black P can cover the infrared spectrum range that is of great interest for applications in medical imaging, night vision and optical communication networks. It is also possible to alloy black P with arsenic to form black $P_xAs_{(1-x)}$ (95). In this alloy, the composition of phosphorus and arsenic may be continuously varied from 0% to 100%, hence potentially tuning the band gap below 0.3 eV and towards 0.1 eV. Such alloys may have band gaps that can cover the spectral range from 5 μm to 12 μm wavelength where many applications of infrared optoelectronics lie, such as high performance thermal imaging and chemical sensing. Furthermore, expertise in chemistry and biology are needed to access the bio-stability and bio-toxicity of this phosphorus-based material. Their compatibility with various biological agents is needed to be accessed for potential electrical and optical sensing applications in biomedical research. This material system also presents challenges and opportunities for the chemists and biologists to work closely together with physicists and engineers in this highly multi-disciplinary field to explore both the fundamentals and applications of this emerging material.

**Figure Captions**

**Fig. 1 Crystal structure and band structure of black P**

(a) Side-view of the black P crystal lattice. The interlayer spacing is 0.53 nm. (b) Top-view (right) of the lattice of single-layer black P. The bond angles are shown. The corresponding x-, y- and z-directions are indicated in both (a) and (b). (c) Band structure of bulk black P mapped out by ARPES measurements. A band gap around 0.3 eV is clearly observed. Superimposed on top are the calculated bands of the bulk crystal. Blue solid and red dashed lines denote empty and filled bands, respectively. The directions of the ARPES mapping are along U (L–Z) and T', as indicated in the first Brillion zone shown in the inset. $E_f$ is the Fermi energy (1). (d) The evolution of the band gap calculated by different methods, and the energy of the optical absorption peak according to the stacking layer number of few-layer black P (31). (c) and (d) are reproduced with permission from ref. (1) and (31) respectively.

**Fig. 2 Electromagnetic wave spectrum and mobility/on-off ratio spectrum**

(a) The electromagnetic wave spectrum and the band gap ranges of various types of 2D materials. The frequency ranges corresponding to the band gaps of 2D materials and their applications in optoelectronics are also indicated (96). (b) the "electronics spectrum", i.e. the mobility/on-off ratio spectrum, of nanomaterials with corresponding performance regions indicated for graphene (97–102) (black squares and grey shaded circle), black P (1–4) (purple dots and light purple shaded circle) and TMDs ($MoS_2$ (103–105), $WSe_2$ (106, 107) and $WS_2$ (108), green dots and light green shaded circle) based transistors. The dots correspond to data from specific references indicated next to them. The shaded regions are the approximate possible ranges of performance reported for the respective materials in the literature.

**Fig. 3 Band gap of thin-film and monolayer black P**

(a) Polarization-resolved infrared relative extinction spectra of a black P thin film when light is polarized along the six directions as shown in the inset. Inset: an optical micrograph of a black P flake with a thickness of around 30 nm. Scale bar: 20 μm. (b) Two representative tunneling spectra in a log scale measured on the black P surface showing a wide band gap with an estimated size of 2.05 eV. Inset: high-resolution STM image ($V_{bias}$=+1.2 V, $I_{set}$=150 pA) with scan size of 2.4 nm × 3.6 nm. (a) and (b) are reproduced with permission from ref.(2) and (81), respectively.

**Fig. 4 Electronic properties of black P thin film**

(a) Sheet conductivity measured as a function of gate voltage for devices with different thicknesses: 10 nm (black solid line), 8 nm (red solid line) and 5 nm (green solid line), with field-effect mobility values of 984, 197 and 55 $cm^2 \cdot V^{-1} \cdot s^{-1}$, respectively. Inset: Field effect mobilities were extracted from the line fit of the linear region of the conductivity (dashed lines). Reproduced with permission from ref. (1). (b) Angle-resolved Hall mobility. Inset: schematic view of a single layer black P showing different crystalline directions. Reproduced with permission from ref. (2). (c) Schematic of the black P transistor device structure. (d) Current and power gain in black P transistors at GHz frequency. The short-circuit current gain $h_{21}$, MSG/MAG, and unilateral power gain U of the 300 nm channel length device after de-embedding. (c) and (d) are reproduced with permission from ref. (58).

**Fig. 5 Anisotropic properties of black P for plasmonics and thermoelectrics applications**

(a) Schematics of black P based plasmonic devices with intrinsic anisotropy in their resonance frequency. The right panel shows the calculated plasmonic dispersions along both the x- and y-directions of a BP crystal. The right panel figure is adopted and modified from ref. (83). (b) Schematics showing the orthogonality between the dominant heat and electron transport directions in single-layer black P, as reported in ref. (69). The figure is inspired by a similar drawing in (69).

**Fig. 6 Protective encapsulation of black P material and device**

(a) AFM images and on-off ratio of black P thin film FETs without and with $AlO_x$ overlayer protection versus ambient exposure time. (a) is reproduced with permission from ref. (53). (b) Schematic and optical micrograph of graphene contacted BP device with boron nitride encapsulation. Red and black dashed areas in the middle panel show the black phosphorus crystal and one of the graphene stripes, respectively. The BN encapsulation layer is also shown. (b) is reproduced with permission from ref. (94).

**Acknowledgments**

The authors thank Prof. David Tomanek for his effort in organizing the first conference solely focused on black phosphorus IPS '14 (Informal Phosphorene Symposium). X.L, S.H and M.S.D acknowledge grant NSF/DMR-1004147 for financial support. H.W. acknowledges support from the Army Research Laboratory. F.X. acknowledges support from the Office of Naval Research (ONR).

**References**

1. Li L, et al. (2014) Black phosphorus field-effect transistors. *Nat Nanotechnol* 9(5):372–7.

2. Xia F, Wang H, Jia Y (2014) Rediscovering black phosphorus as an anisotropic layered material for optoelectronics and electronics. *Nat Commun* 5:4458.

3. Liu H, et al. (2014) Phosphorene: An Unexplored 2D Semiconductor with a High Hole Mobility. *ACS Nano* 8(4):4033–4041.

4. Koenig SP, Doganov RA, Schmidt H, Neto AHC, Özyilmaz B (2014) Electric field effect in ultrathin black phosphorus. *Appl Phys Lett* 104(10):103106.

5. Rodin AS, Carvalho A, Castro Neto AH (2014) Strain-Induced Gap Modification in Black Phosphorus. *Phys Rev Lett* 112(17):176801.

6. Qiao J, Kong X, Hu Z-X, Yang F, Ji W (2014) High-mobility transport anisotropy and linear dichroism in few-layer black phosphorus. *Nat Commun* 5:4475.

7. Churchill HOH, Jarillo-Herrero P (2014) Two-dimensional crystals: Phosphorus joins the family. *Nat Nanotechnol* 9(5):330–331.

8. Liu H, Du Y, Deng Y, Ye PD (2014) Semiconducting black phosphorus: synthesis, transport properties and electronic applications. *Chem Soc Rev*:online.

9. Clark SM, Zaug JM (2010) Compressibility of cubic white, orthorhombic black, rhombohedral black, and simple cubic black phosphorus. *Phys Rev B* 82(13):134111.

10. Morita A (1986) Semiconducting black phosphorus. *Appl Phys A Solids Surfaces* 39(4):227–242.

11. Asahina H, Shindo K, Morita A (1982) Electronic Structure of Black Phosphorus in Self-Consistent Pseudopotential Approach. *J Phys Soc Japan* 51(4):1193–1199.

12. Takao Y, Morita A (1981) Electronic structure of black phosphorus: Tight binding approach. *Phys B+C* 105(1-3):93–98.

13. Bridgman PW (1914) Two New Modifications of Phosphorus. *J Am Chem Soc* 36(7):1344–1363.

14. Jamieson JC (1963) Crystal Structures Adopted by Black Phosphorus at High Pressures. *Science (80- )* 139(3561):1291–1292.

15. Brown A, Rundqvist S (1965) Refinement of the crystal structure of black phosphorus. *Acta Crystallogr* 19(4):684–685.

16. Akahama Y, Endo S, Narita S (1983) Electrical Properties of Black Phosphorus Single Crystals. *J Phys Soc Japan* 52(6):2148–2155.

17. Ikezawa M, Kondo Y, Shirotani I (1983) Infrared Optical Absorption Due to One and Two Phonon Processes in Black Phosphorus. *J Phys Soc Japan* 52(5):1518–1520.

18. Suzuki N, Aoki M (1987) Interplanar forces of black phosphorus caused by electron-lattice interaction. *Solid State Commun* 61(10):595–600.

19. Kawamura H, Shirotani I, Tachikawa K (1985) Anomalous superconductivity and pressure induced phase transitions in black phosphorus. *Solid State Commun* 54(9):775–778.

20. Wittig J, Matthias BT (1968) Superconducting Phosphorus. *Science (80- )* 160(3831):994–995.

21. Park C-M, Sohn H-J (2007) Black Phosphorus and its Composite for Lithium Rechargeable Batteries. *Adv Mater* 19(18):2465–2468.

22. Nagao M, Hayashi A, Tatsumisago M (2011) All-solid-state lithium secondary batteries with high capacity using black phosphorus negative electrode. *J Power Sources* 196(16):6902–6905.

23. Sun J, et al. (2014) Formation of Stable Phosphorus–Carbon Bond for Enhanced Performance in Black Phosphorus Nanoparticle–Graphite Composite Battery Anodes. *Nano Lett* 14(8):4573–4580.

24. Keyes R (1953) The Electrical Properties of Black Phosphorus. *Phys Rev* 92(3):580–584.

25. Morita A, Sasaki T (1989) Electron-Phonon Interaction and Anisotropic Mobility in Black Phosphorus. *J Phys Soc Japan* 58(5):1694–1704.

26. Asahina H, Morita A (1984) Band structure and optical properties of black phosphorus. *J Phys C Solid State Phys* 17(11):1839–1852.

27. Sugai S, Shirotani I (1985) Raman and infrared reflection spectroscopy in black phosphorus. *Solid State Commun* 53(9):753–755.

28. Shibata K, Sasaki T, Morita A (1987) The Energy Band Structure of Black Phosphorus and Angle-Resolved Ultraviolet Photoelectron Spectra. *J Phys Soc Japan* 56(6):1928–1931.

29. Narita S, et al. (1983) Far-Infrared Cyclotron Resonance Absorptions in Black Phosphorus Single Crystals. *J Phys Soc Japan* 52(10):3544–3553.

30. Baba M, Izumida F, Morita A, Koike Y, Fukase T (1991) Electrical Properties of Black Phosphorus Single Crystals Prepared by the Bismuth-Flux Method. *Jpn J Appl Phys* 30(Part 1, No. 8):1753–1758.

31. Tran V, Soklaski R, Liang Y, Yang L (2014) Layer-controlled band gap and anisotropic excitons in few-layer black phosphorus. *Phys Rev B* 89(23):235319.

32. Rudenko AN, Katsnelson MI (2014) Quasiparticle band structure and tight-binding model for single- and bilayer black phosphorus. *Phys Rev B* 89(20):201408.

33. Kim H (2014) Effect of van der Waals interaction on the structural and cohesive properties of black phosphorus. *J Korean Phys Soc* 64(4):547–553.

34. Li Y, Yang S, Li J (2014) Modulation of the Electronic Properties of Ultrathin Black Phosphorus by Strain and Electrical Field. *J Phys Chem C* 118(41):23970–23976.

35. Fei R, Yang L (2014) Strain-Engineering the Anisotropic Electrical Conductance of Few-Layer Black Phosphorus. *Nano Lett* 14(5):2884–2889.

36. Wei Q, Peng X (2014) Superior mechanical flexibility of phosphorene and few-layer black phosphorus. *Appl Phys Lett* 104(25):251915.

37. Fei R, Yang L (2014) Lattice vibrational modes and Raman scattering spectra of strained phosphorene. *Appl Phys Lett* 105(8):083120.

38. Peng X, Wei Q, Copple A (2014) Strain-engineered direct-indirect band gap transition and its mechanism in two-dimensional phosphorene. *Phys Rev B* 90(8):085402.

39. Liu Y, Xu F, Zhang Z, Penev ES, Yakobson BI (2014) Two-dimensional mono-elemental semiconductor with electronically inactive defects: the case of phosphorus. *Nano Lett* 14(12):6782–6.

40. Yu X, Ushiyama H, Yamashita K (2014) Comparative study of Na/Li intercalation and diffusion mechanism in black phosphorus from first-principles simulation. *Chem Lett*:140741.

41. Zhu Z, Tománek D (2014) Semiconducting Layered Blue Phosphorus: A Computational Study. *Phys Rev Lett* 112(17):176802.

42. Guan J, Zhu Z, Tománek D (2014) Phase Coexistence and Metal-Insulator Transition in Few-Layer Phosphorene: A Computational Study. *Phys Rev Lett* 113(4):046804.

43. Ramasubramaniam A, Muniz AR (2014) Ab initio studies of thermodynamic and electronic properties of phosphorene nanoribbons. *Phys Rev B* 90(8):085424.

44. Tran V, Yang L (2014) Scaling laws for the band gap and optical response of phosphorene nanoribbons. *Phys Rev B* 89(24):245407.

45. Guo H, Lu N, Dai J, Wu X, Zeng XC (2014) Phosphorene Nanoribbons, Phosphorus Nanotubes, and van der Waals Multilayers. *J Phys Chem C* 118(25):14051–14059.

46. Peng X, Copple A, Wei Q (2014) Edge effects on the electronic properties of phosphorene nanoribbons. *J Appl Phys* 116(14):144301.

47. Han X, Morgan Stewart H, Shevlin SA, Catlow CRA, Guo ZX (2014) Strain and Orientation Modulated Bandgaps and Effective Masses of Phosphorene Nanoribbons. *Nano Lett* 14(8):4607–4614.

48. Dai J, Zeng XC (2014) Bilayer Phosphorene: Effect of Stacking Order on Bandgap and Its Potential Applications in Thin-Film Solar Cells. *J Phys Chem Lett* 5(7):1289–1293.

49. Castellanos-Gomez A, et al. (2014) Isolation and characterization of few-layer black phosphorus. *2D Mater* 1(2):025001.

50. Zhang S, et al. (2014) Extraordinary Photoluminescence and Strong Temperature/Angle-Dependent Raman Responses in Few-Layer Phosphorene. *ACS Nano* 8(9):9590–9596.

51. Lu W, et al. (2014) Plasma-assisted fabrication of monolayer phosphorene and its Raman characterization. *Nano Res* 7(6):853–859.

52. Dai J, Zeng XC (2014) Structure and stability of two dimensional phosphorene with =O or =NH functionalization. *RSC Adv* 4(89):48017–48021.

53. Wood JD, et al. (2014) Effective Passivation of Exfoliated Black Phosphorus Transistors against Ambient Degradation. *Nano Lett*: 14 (12):6964–6970.

54. Favron A, et al. Exfoliating pristine black phosphorus down to the monolayer: photo-oxidation and electronic confinement effects. *arXiv:14080345v2*.

55. Na J, et al. (2014) Few-layer black phosphorus field-effect transistors with reduced current fluctuation. *ACS Nano* 8(11):11753–62.

56. Jiang J-W, Park HS (2014) Negative poisson's ratio in single-layer black phosphorus. *Nat Commun* 5:4727.

57. Jiang J-W, Park HS (2014) Mechanical properties of single-layer black phosphorus. *J Phys D Appl Phys* 47(38):385304.

58. Wang H, et al. (2014) Black Phosphorus Radio-Frequency Transistors. *Nano Lett* 14(11):6424–6429.

59. Du Y, Liu H, Deng Y, Ye PD (2014) Device Perspective for Black Phosphorus Field-Effect Transistors: Contact Resistance, Ambipolar Behavior, and Scaling. *ACS Nano* 8(10):10035–10042.

60. The Effect of Dielectric Capping on Few-Layer Phosphorene Transistors: Tuning the Schottky Barrier Heights (2014) *IEEE Electron Device Lett* 35(7):795–797.

61. Das S, et al. (2014) Tunable Transport Gap in Phosphorene. *Nano Lett* 14(10):5733–5739.

62. Engel M, Steiner M, Avouris P (2014) A black phosphorus photo-detector for multispectral, high-resolution imaging. *arXiv:14072534*.

63. Low T, Engel M, Steiner M, Avouris P (2014) Origin of photoresponse in black phosphorus phototransistors. *Phys Rev B* 90(8):081408.

64. Buscema M, et al. (2014) Fast and Broadband Photoresponse of Few-Layer Black Phosphorus Field-Effect Transistors. *Nano Lett* 14(6):3347–3352.

65. Deng Y, et al. (2014) Black Phosphorus–Monolayer $MoS_2$ van der Waals Heterojunction p–n Diode. *ACS Nano* 8(8):8292–8299.

66. Buscema M, Groenendijk DJ, Steele GA, van der Zant HSJ, Castellanos-Gomez A (2014) Photovoltaic effect in few-layer black phosphorus PN junctions defined by local electrostatic gating. *Nat Commun* 5:4651.

67. Hong T, et al. (2014) Polarized photocurrent response in black phosphorus field-effect transistors. *Nanoscale* 6(15):8978.

68. Qin G, et al. (2014) Hinge-like structure induced unusual properties of black phosphorus and new strategies to improve the thermoelectric performance. *Sci Rep* 4:6946.

69. Fei R, et al. (2014) Enhanced Thermoelectric Efficiency via Orthogonal Electrical and Thermal Conductances in Phosphorene. *Nano Lett.*, 14 (11):6393–6399

70. Kou L, Frauenheim T, Chen C (2014) Phosphorene as a Superior Gas Sensor: Selective Adsorption and Distinct $I-V$ Response. *J Phys Chem Lett* 5(15):2675–2681.

71. Wang X, et al. Highly Anisotropic and Robust Excitons in Monolayer Black Phosphorus. *arXiv:14111695v1*.

72. Ezawa M (2014) Topological origin of quasi-flat edge band in phosphorene. *New J Phys* 16(11):115004.

73. Wu Q, et al. Band Gaps and Giant Stark Effect in Nonchiral Phosphorene Nanoribbons. *arXiv:14053077v3*.

74. Novoselov KS, et al. (2005) Two-dimensional gas of massless Dirac fermions in graphene. *Nature* 438(7065):197–200.

75. Zhang Y, Tan Y-W, Stormer HL, Kim P (2005) Experimental observation of the quantum Hall effect and Berry's phase in graphene. *Nature* 438(7065):201–204.

76. Mak KF, Lee C, Hone J, Shan J, Heinz TF (2010) Atomically Thin MoS2: A New Direct-Gap Semiconductor. *Phys Rev Lett* 105(13):136805.

77. Splendiani A, et al. (2010) Emerging Photoluminescence in Monolayer $MoS_2$. *Nano Lett* 10(4):1271–1275.

78. Jones AM, et al. (2014) Spin–layer locking effects in optical orientation of exciton spin in bilayer WSe2. *Nat Phys* 10(2):130–134.

79. Britnell L, et al. (2013) Strong light-matter interactions in heterostructures of atomically thin films. *Science* 340(6138):1311–4.

80. Low T, et al. (2014) Tunable optical properties of multilayer black phosphorus thin films. *Phys Rev B* 90(7):075434.

81. Liang L, et al. (2014) Electronic Bandgap and Edge Reconstruction in Phosphorene Materials. *Nano Lett* 14(11):6400–6406.

82. Gong K, Zhang L, Ji W, Guo H (2014) Electrical contacts to monolayer black phosphorus: A first-principles investigation. *Phys Rev B* 90(12):125441.

83. Low T, et al. (2014) Plasmons and Screening in Monolayer and Multilayer Black Phosphorus. *Phys Rev Lett* 113(10):106802.

84. Lv H, Lu W, Shao D, Sun Y Large thermoelectric power factors in black phosphorus and phosphorene. *arXiv:14045171v1*.

85. Zhang J, et al. (2014) Phosphorene nanoribbon as a promising candidate for thermoelectric applications. *Sci Rep* 4:6452.

86. Zhang, S., Yang, J., Xu, R. J., Wang, F., Li, W. F., Ghufran, M., Zhang, Y. W., Yu, Z. F., Zhang, G., Qin, Q. H., Lu, Y. H. (2014) Extrordinary PL and strong temperature angle-dependent Raman responses in few-layer phosphorene. *arXiv:14070502*.

87. Rissi EN, Soignard E, McKiernan KA, Benmore CJ, Yarger JL (2012) Pressure-induced crystallization of amorphous red phosphorus. *Solid State Commun* 152(5):390–394.

88. Endo S, Akahama Y, Terada S, Narita S (1982) Growth of Large Single Crystals of Black Phosphorus under High Pressure. *Jpn J Appl Phys* 21(Part 2, No. 8):L482–L484.

89. Krebs H, Weitz H, Worms KH (1955) Die katalytische Darstellung des schwarzen Phosphors. *Zeitschrift für Anorg und Allg Chemie* 280(1-3):119–133.

90. Lange S, Schmidt P, Nilges T (2007) $Au_3SnP_7$@Black Phosphorus: An Easy Access to Black Phosphorus. *Inorg Chem* 46(10):4028–4035.

91. Nilges T, Kersting M, Pfeifer T (2008) A fast low-pressure transport route to large black phosphorus single crystals. *J Solid State Chem* 181(8):1707–1711.

92. Köpf M, et al. (2014) Access and in situ growth of phosphorene-precursor black phosphorus. *J Cryst Growth* 405:6–10.

93. Luo X, et al. Temporal and Thermal Stability of Al2O3-passivated Phosphorene MOSFETs. *arXiv:14100994v1*.

94. Avsar A, et al. (2014) Electrical characterization of fully encapsulated ultra thin black phosphorus-based heterostructures with graphene contacts. *Arxiv: 14121191*.

95. Osters O, et al. (2012) Synthesis and Identification of Metastable Compounds: Black Arsenic-Science or Fiction? *Angew Chemie Int Ed* 51(12):2994–2997.

96. Xia F, Wang H, Xiao D, Dubey M, Ramasubramaniam A Two-Dimensional Material Nanophotonics. *Nat Photonics* 8(12):899–907.

97. Liang X, Fu Z, Chou SY (2007) Graphene Transistors Fabricated via Transfer-Printing In Device Active-Areas on Large Wafer. *Nano Lett* 7(12):3840–3844.

98. Lemme MC, Echtermeyer TJ, Baus M, Kurz H (2007) A Graphene Field-Effect Device. *IEEE Electron Device Lett* 28(4):282–284.

99. Das A, et al. (2008) Monitoring dopants by Raman scattering in an electrochemically top-gated graphene transistor. *Nat Nanotechnol* 3(4):210–5.

100. Kedzierski J, et al. (2008) Epitaxial Graphene Transistors on SiC Substrates. *IEEE Trans Electron Devices* 55(8):2078–2085.

101. Park J, et al. (2012) Single-Gate Bandgap Opening of Bilayer Graphene by Dual Molecular Doping. *Adv Mater* 24(3):407–411.

102. Szafranek BN, Schall D, Otto M, Neumaier D, Kurz H (2011) High On/Off Ratios in Bilayer Graphene Field Effect Transistors Realized by Surface Dopants. *Nano Lett* 11(7):2640–2643.

103. Ayari A, Cobas E, Ogundadegbe O, Fuhrer MS (2007) Realization and electrical characterization of ultrathin crystals of layered transition-metal dichalcogenides. *J Appl Phys* 101(1):014507.

104. Kim S, et al. (2012) High-mobility and low-power thin-film transistors based on multilayer MoS2 crystals. *Nat Commun* 3:1011.

105. Wu W, et al. (2013) High mobility and high on/off ratio field-effect transistors based on chemical vapor deposited single-crystal MoS2 grains. *Appl Phys Lett* 102(14):142106.

106. Fang H, et al. (2012) High-Performance Single Layered $WSe_2$ p-FETs with Chemically Doped Contacts. *Nano Lett* 12(7):3788–3792.

107. Liu W, et al. (2013) Role of Metal Contacts in Designing High-Performance Monolayer n-Type $WSe_2$ Field Effect Transistors. *Nano Lett* 13(5):1983–1990.

108. Ovchinnikov D, Allain A, Huang Y-S, Dumcenco D, Kis A (2014) Electrical Transport Properties of Single-Layer $WS_2$. *ACS Nano* 8(8):8174–8181.

# Figure 1. Crystal structure of black phosphorus

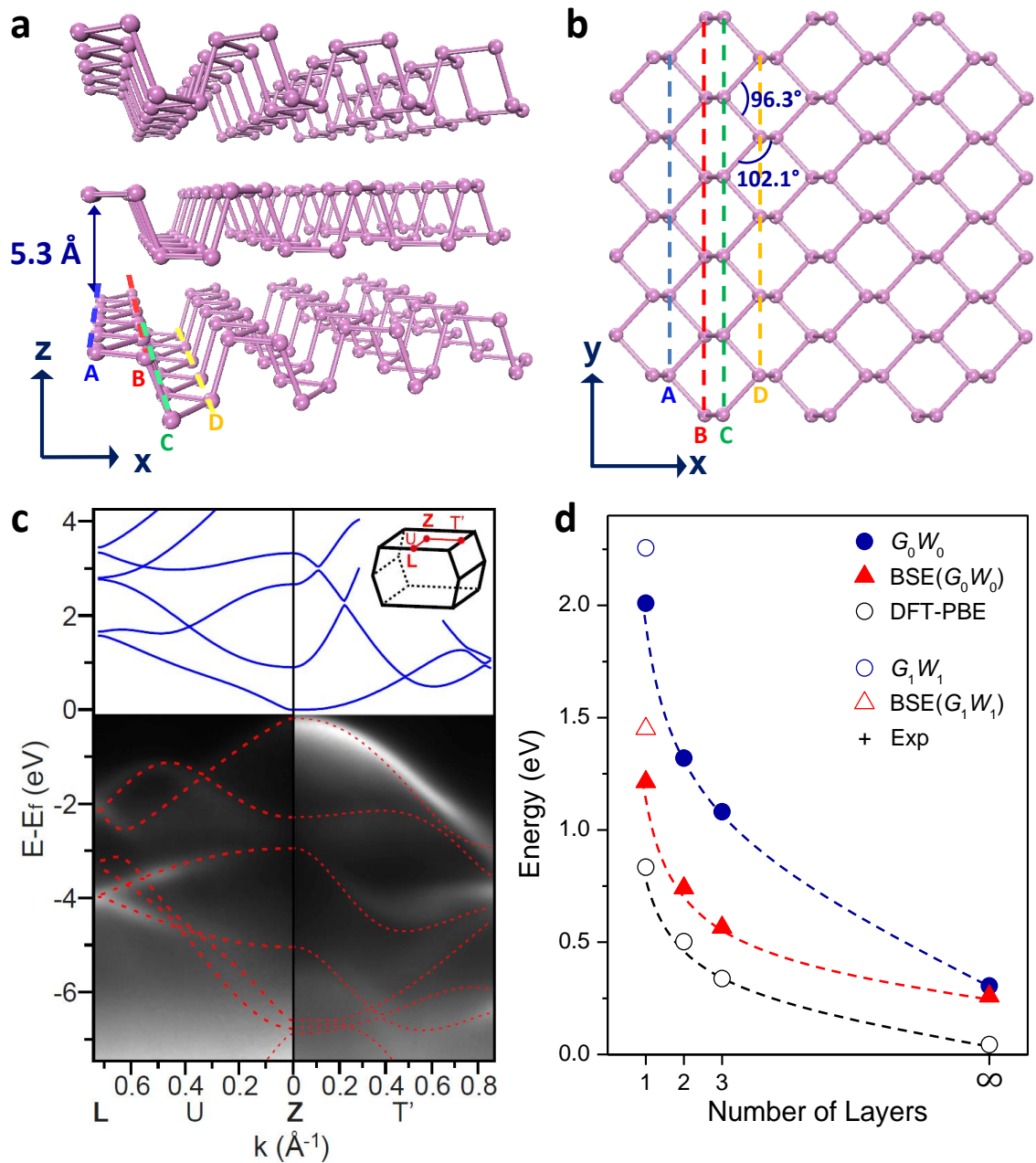

**Figure 2. Electromagnetic wave spectrum and Mobility/on-off ratio spectrum**

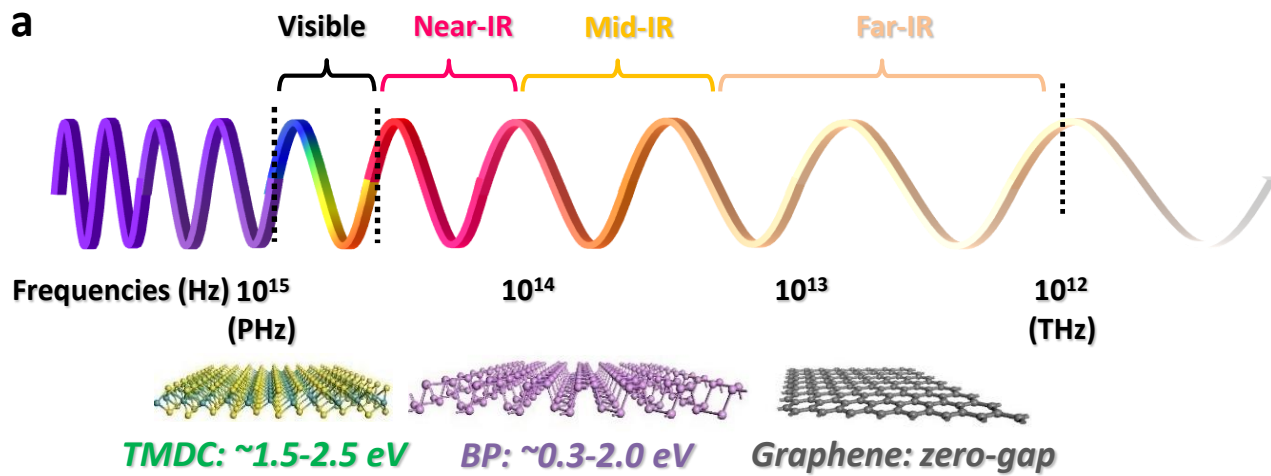

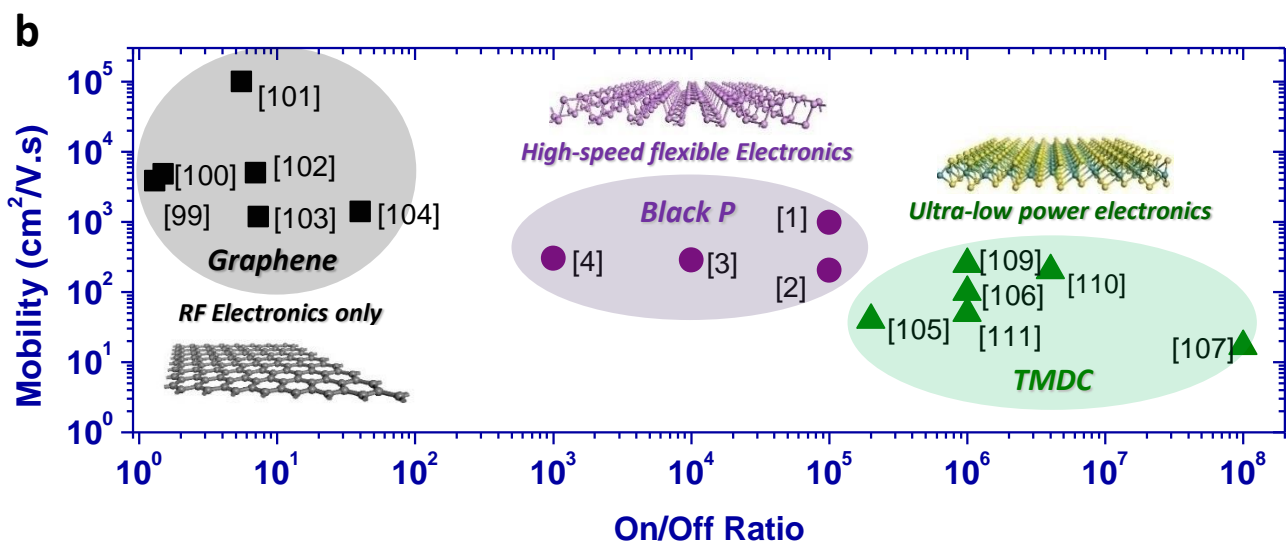

# Figure 3. Band gap of thin-film and monolayer black phosphorus

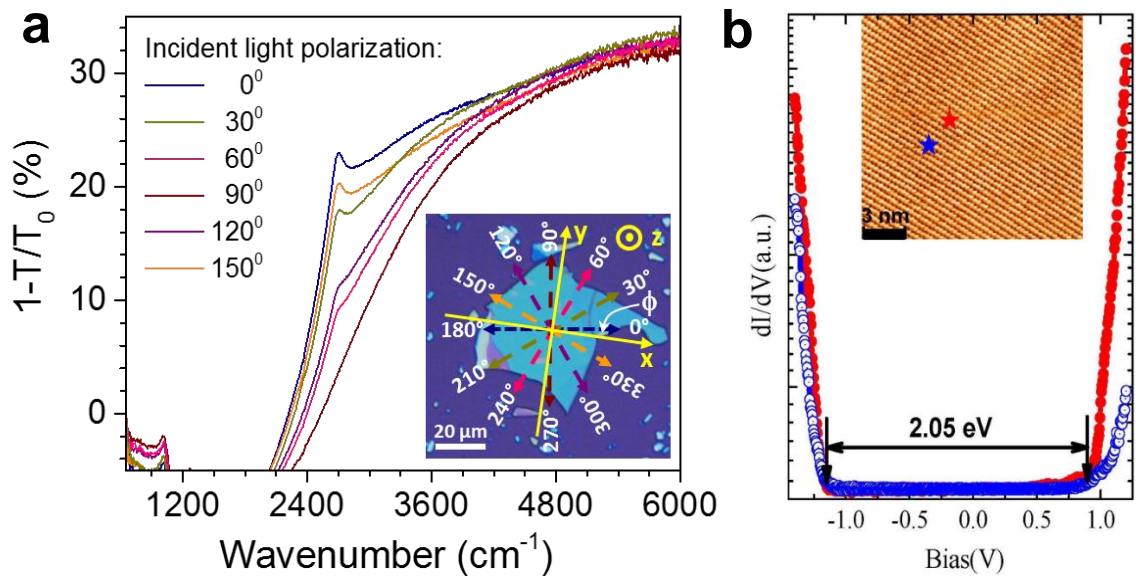

# Figure 4. Electronic properties of black phosphorus thin film

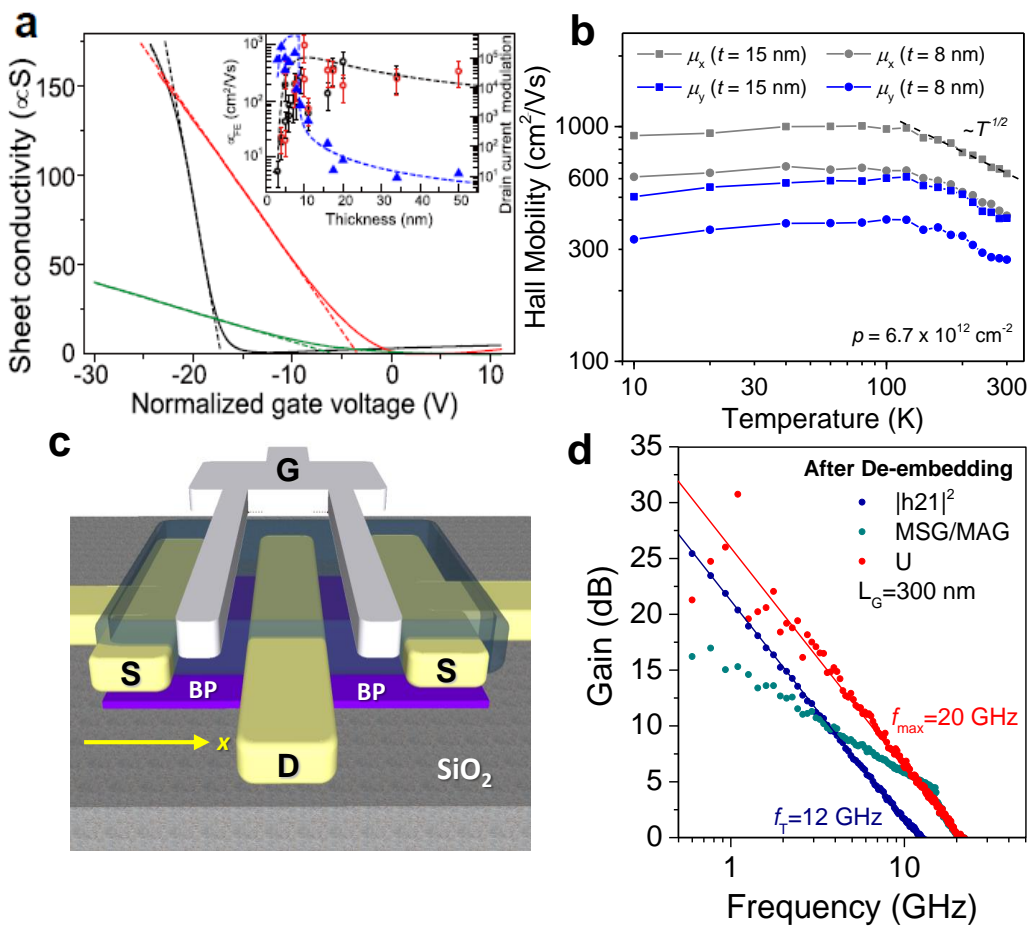

# Figure 5. Anisotropic Properties of BP for plasmonics and thermoelectrics applications

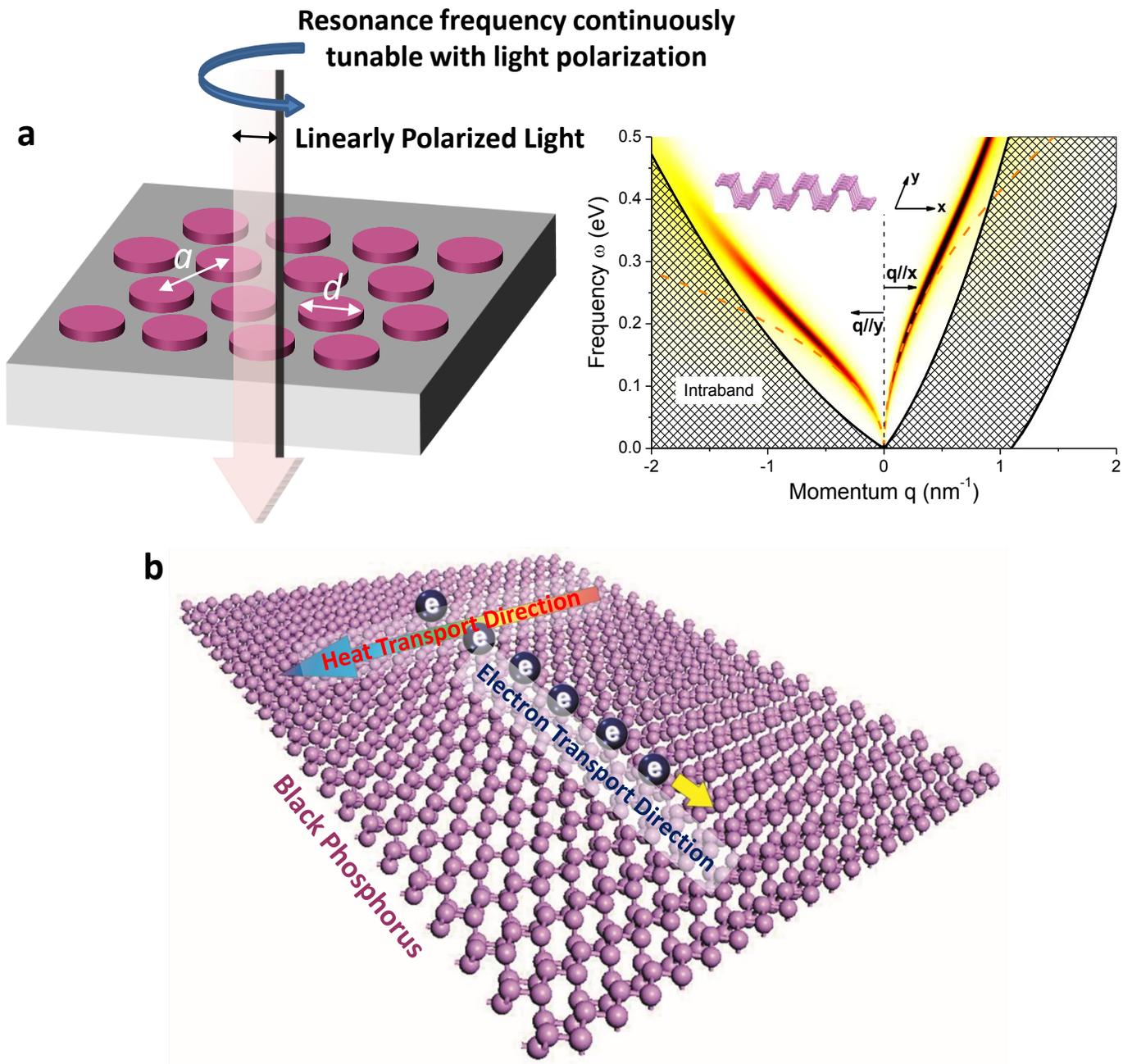

# Figure 6. Protective encapsulation of black P material and device

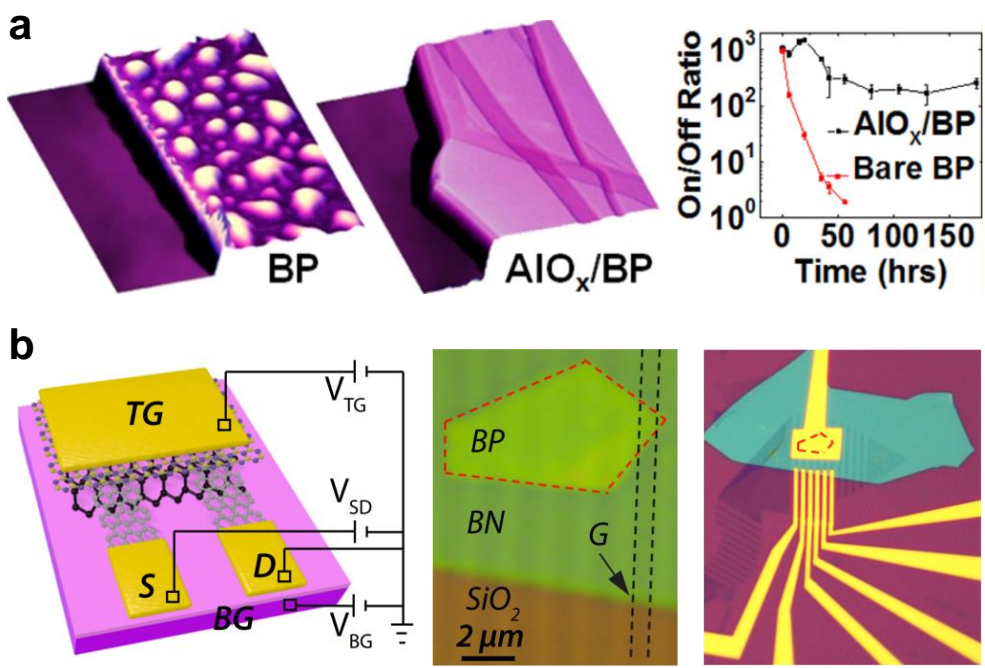